\documentclass[%
 reprint,
nofootinbib,
 amsmath,amssymb,
 aps,
showkeys
]{revtex4-2}

\UseRawInputEncoding

\usepackage{array}
\newcolumntype{L}[1]{>{\raggedright\let\newline\\\arraybackslash\hspace{0pt}}m{#1}}
\newcolumntype{C}[1]{>{\centering\let\newline\\\arraybackslash\hspace{0pt}}m{#1}}
\newcolumntype{R}[1]{>{\raggedleft\let\newline\\\arraybackslash\hspace{0pt}}m{#1}}

\usepackage{float}
\usepackage{graphicx}
\graphicspath{ {./images/} }
\usepackage{dcolumn}
\usepackage{bm}
\usepackage{hyperref}
\usepackage{xcolor}

\hypersetup{
    colorlinks=true,
    linkcolor=blue,
    filecolor=magenta,      
    urlcolor=blue,
}

\begin{document}

\preprint{APS/123-QED}

\title{A fully-automated end-to-end pipeline for \\ massive black hole binary signal extraction from LISA data}

\author{Michael L. Katz}
\email{michael.katz@aei.mpg.de}
\affiliation{Max-Planck-Institut f\"ur Gravitationsphysik, Albert-Einstein-Institut, 
Am M\"uhlenberg 1, 14476 Potsdam-Golm, Germany}

 \begin{abstract}
 
 The LISA Data Challenges Working Group within the LISA Consortium has started publishing datasets to benchmark, compare, and build LISA data analysis infrastructure as the Consortium prepares for the launch of the mission. We present our solution to the dataset from LISA Data Challenge (LDC) 1A containing a single massive black hole binary signal. This solution is built from a fully-automated and GPU-accelerated pipeline consisting of three segments: a brute-force initial search; a refining search that uses the efficient Likelihood computation technique of Heterodyning (also called Relative Binning) to locate the maximum Likelihood point; and a parameter estimation portion that also takes advantage of the speed of the Heterodyning method. This pipeline takes tens of minutes to evolve from randomized initial parameters throughout the prior volume to a converged final posterior distribution. Final posteriors are shown for both datasets from LDC 1A: one noiseless data stream and one containing additive noise. A posterior distribution including higher harmonics is also shown for a self-injected waveform with the same source parameters as is used in the original LDC 1A dataset. This higher-mode posterior is shown in order to provide a more realistic distribution on the parameters of the source.

\end{abstract}

      
\maketitle


\section{Introduction}\label{sec:intro}

The era of gravitational-wave physics and astronomy is here in full force. Detections of $\sim10$s of binaries by ground-based observing groups \cite{LIGOScientific:2018mvr, LVK2018LivingReview} have shown the power of this new messenger for understanding the Universe. The future space-based gravitational wave detector, the Laser Interferometer Space Antenna (LISA) \cite{LISAMissionProposal}, will open a new window into the milliHertz regime of the gravitational-wave spectrum. This frequency range contains many sources of interest including massive black hole binaries (MBHB) stemming from the mergers of their host galaxies \cite[e.g.][]{Begelman1980}; stellar-origin black hole binaries (SOBHB) earlier in their evolution that can act as progenitors to ground-based sources \cite[e.g.][]{Sesana:2016ljz}; extreme-mass-ratio-inspirals (EMRI) involving a compact object bound to an MBH in a close and energetic orbit \cite[e.g.][]{Babak:2017tow}; and Galactic binaries (GBs) typically consisting of two white dwarf stars in a bound system at large separations exhibiting a quasimonochromatic signal \cite[e.g.][]{Littenberg:2020bxy}. 

MBHBs are a primary focus of the LISA mission due to their unique capabilities in helping to answer many scientific questions in astrophysics, cosmology, and fundamental physics \cite{Schutz1986, Holz2005, Petiteau2011, Burke-Spolaor2013, Gair2013, Barausse2015, Bogdanovic2015, Klein:2015hvg, LISAMissionProposal}. This includes probing the mass spectrum of MBHs and their formation channels, as well as better understanding their interplay within the formation and evolution of galaxies. These systems with total detector-frame masses of $\sim10^5-10^7 M_\odot$ are expected to have high signal-to-noise ratios (SNR), of order $\sim100-1000$s, allowing for in-depth extraction of parameters and detailed tests of General Relativity. Recent event rate estimates for MBHBs detectable by LISA range from $\sim1-20$ per year, depending on the underlying assumptions \cite{Klein:2015hvg, Berti2016, Salcido2016, Katz:2019qlu, Bonetti2019}. 

While there is a large amount of science that can be done by observing these MBHB systems, properly analyzing the LISA data is not an easy task. All of the sources (including all other source classes) will be emitting signals that will overlap throughout the lifetime of the LISA mission, both requiring the use of global fitting of source parameters and preventing direct access to pure signal-free noise information as is available for current ground-based observing runs. Due to the long-lived signals, the time-dependence of the detector orientation will need to be included in the analysis to ensure unbiased extraction of source parameters. There is expected to be non-stationary noise effects in the form of data gaps and glitches, as well as expected drift of the overall instrument sensitivity over its observing lifetime. The new LISA Data Challenges (LDC) Working Group is tasked with slowly incorporating all of these effects and building the initial data analysis pipelines that will, over time, prove the capability of the LISA community to properly extract source parameters and scientific information from the future realistic LISA data stream. The LDC provides datasets that help researchers to compare and contrast methods and results for the same underlying data, allowing the community to more fully understand its range of capabilities. 

Earlier work over the past few decades to address the analysis of MBHBs with LISA can be found in \cite{Cutler1994, Vecchio2004, Berti2005, Arun2006, Lang2006, Thorpe2009, McWilliams2010, McWilliams2010b, McWilliams2011, Arun2007, Trias2008, Porter2008, McWilliams2010, Baibhav:2020tma, Brown2007, Cornish2006, Crowder2006, Wickham2006, Rover2007, Feroz2009, Gair2009, Petiteau2009, Babak2010, Porter2014, Porter2015}. More recent analyses in \cite{Marsat:2020rtl} and \cite{Katz:2020hku} were performed with a more modern LISA sensitivity \cite{LISAMissionProposal, SciRD1}; full inspiral-merger-ringdown waveform models that included higher harmonic modes; and a more accurate representation of the LISA response from \cite{Marsat:2018oam}. Waveforms in \cite{Katz:2020hku} also contained aligned spins. The effect of data gaps on MBHB analysis was recently looked at without noise in \cite{Dey:2021dem}. There have also been papers addressing the simultaneous analysis of MBHBs with LISA and other space-based detectors \cite[e.g.][]{Ruan:2019tje, Shuman:2021ruh, Zhang:2021kkh}. 

The first round of the new LDC datasets (LDC-1A) focuses on single sources injected into stationary instrumental LISA noise. \textit{This work will present a solution for the LDC-1A dataset containing a single MBHB source. Specifically, we will describe a fully-automated end-to-end pipeline capable of extracting accurate posterior distributions on source parameters in under one hour of runtime with no human intervention or prior knowledge of the signal in the data}. Authors in \cite{Cornish:2020vtw} presented a full pipeline for the search and parameter estimation of MBHBs in response to the LDC-1A dataset. This was recently extended to the MBHB portion of the new LDC-2A challenge \cite{Cornish:2021smq}. This work is similar in its origins and goals to our work presented here. The duration of the pipelines between the two works are similar. The parameter estimation stage of the pipelines are both run with fast, residual-based Likelihood methods, but the search stages are different. In \cite{Cornish:2020vtw} , the search portion relies on a sequence of marginalizations over certain parameters and a fixed LISA detector. This setup improves the speed of the search, but it requires updating when more physics are included (e.g. precession). In our work, we achieve a fast search by using computational acceleration from Graphics Processing Units (GPUs), rather than marginalization. Our search uses a brute-force approach in the initial stages, which should extend to more complicated waveform models, as well as searching for sources in their early inspiral where movement of the LISA detector will be required for the earliest possible detection. We follow this initial stage with a maximum Likelihood refining step that makes use of the very efficient Likelihood method of Heterodyning \cite{Cornish:2010kf, Cornish:2021lje}. We use the specific Heterodyning formulation given in \cite{Zackay:2018qdy}. In \cite{Zackay:2018qdy}, they refer to their Heterodyning implementation as ``Relative Binning.'' 

In addition to providing a solution to the LDC datasets, we also provide a posterior distribution with a waveform containing higher harmonic modes for the same source parameters as the signal in the LDC data. The LDC data only includes the dominant $l=m=2$ harmonic. The higher-harmonic posterior is provided to give the reader a better understanding of the constraints on parameters and the posterior distribution in general when higher modes are included in the waveform. MBHBs detectable with LISA are expected to have multiple detectable harmonics given the high SNR of these sources.

We expect the methods presented here to extend to datasets with multiple MBHBs. This assumes the predicted rates for MBHB detections are roughly correct, meaning the merger-ringdown portion of coalescence, where most of the SNR is accumulated for MBHBs, is unlikely to overlap with another signal's merger-ringdown since this portion of the signal has a short duration of $\sim$hour or less. MBHBs that have overlapping merger-ringdown signals will require more sophisticated methods such as global fitting. 

In Section~\ref{sec:bayes_mbhb} we briefly describe the waveform, LISA response function, and Likelihood computations used in this work. The pipeline that uses these methods and the final posteriors it produces are provided in Section~\ref{sec:pipeline}. We then discuss the pipeline in the larger context of LISA data analysis in Section~\ref{sec:discuss} and conclude our remarks in Section~\ref{sec:conclusion}.

\section{Bayesian methods for massive black hole binary parameter extraction with LISA}\label{sec:bayes_mbhb}

\begin{figure}[t]
\begin{center}
\includegraphics[scale=0.55]{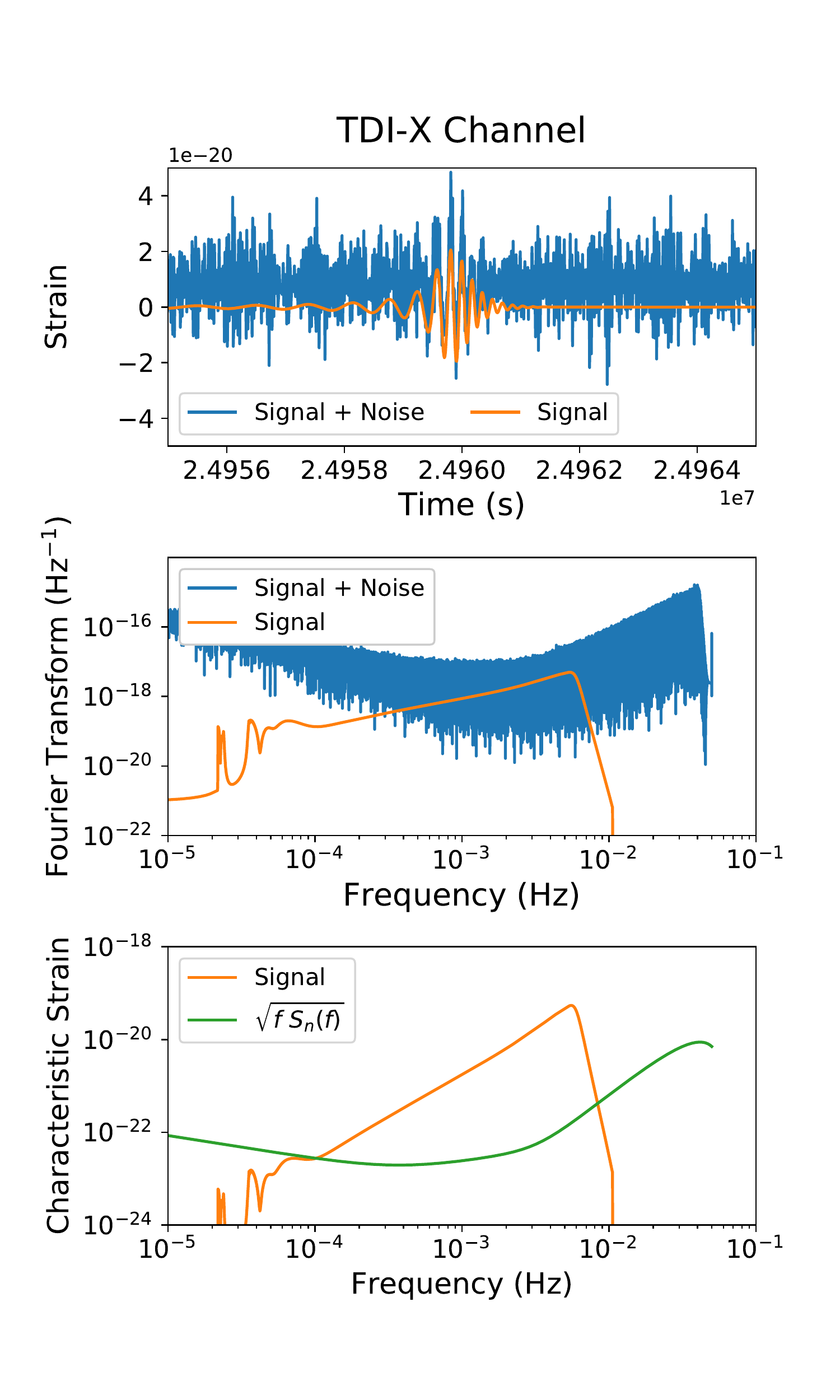}
\caption{The data from the LDC-1A dataset containing one MBHB signal. The top panel shows the time domain representation of the TDI X-channel (see Section~\ref{sec:template_gen}), zooming in around the merger-ringdown. The middle panel is the corresponding frequency domain representation shown as the Fourier Transform of the data. The dataset without noise is shown in orange while the dataset containing noise is shown in blue. The bottom panel shows the noiseless signal (orange) in the characteristic strain representation: $h_c^2 = 4\left| f^2\tilde{h}(f)^2 \right|$. The sensitivity curve (green) is also shown as the characteristic strain in the noise: $h_n^2 = fS_n(f)$ \cite{Finn2000, Moore2015}.}\label{fig:signal_example}
\end{center}
\end{figure}

The gravitational-wave data stream, $d(t)$, is usually represented as a sum of noise and signal contributions: $d(t)=s(t) + n(t)$, where $s(t)$ is the true underlying gravitational-wave signal and $n(t)$ is the noise contribution. In our case, $d(t)$ is provided by the LDC dataset. The data from the LDC are shown in both the time and frequency domain in Figure~\ref{fig:signal_example}. We will analyze two datasets with the same $s(t)$ injection, but with different $n(t)$. In one case, $n(t)=0$ for all values of $t$, which we will refer to as the ``noiseless'' case. In the ``noisy'' or ``noise-infused'' dataset, $n(t)$, which is modeled as a stochastic process, is generated by the LDC Working Group. The noise curve used is the ``SciRDv1'' noise curve from the LDC code base \cite{SciRD1}. The Fourier transform of the noise is represented as $\tilde{n}(f) = \mathcal{F}\{n(t)\}$. Because the noise in the time domain is assumed to be Gaussian and stationary, the noise in the frequency domain is Gaussian with mean zero and a covariance matrix that is diagonal with values given by the power spectral density (PSD) in the noise: $\langle \tilde{n}(f) \tilde{n}(f')\rangle = \frac{1}{2}S_n(f)\delta(f - f')$ (with $f\geq0$), where $S_n(f)$ represents the power in the noise at a given frequency.

The goal of parameter estimation is to determine the posterior distribution on the parameters that describe the underlying signal given the specific set of noisy data and the modeling of the data generation process. This conditional probability can be determined using Bayes' Theorem:
\begin{equation}\label{eq:bayes}
	p(\vec{\Theta}|d, \Lambda) = \frac{p(d | \vec{\Theta}, \Lambda)p(\vec{\Theta}|\Lambda)}{p(d |\Lambda)},
\end{equation}
where $\vec{\Theta}$ is the vector of parameters that describe the signal $s(t)$ and $\Lambda$ is the chosen underlying model. In reality, $\Lambda$ will be an approximation to the true signal given by nature. However, in this work, we assume that $s(t)$ is exactly given by $\Lambda$ and avoid assessing any biases due to incorrect modeling. We will, therefore, drop $\Lambda$ from the notation for the remainder of this paper. See \cite{Husa2016, Khan2016, London2018} for information on waveform modeling errors related to the waveforms used in this paper.  

On the right-hand side of Equation~\ref{eq:bayes}, $p(\vec{\Theta})$ represents the prior probability distribution of the parameters that describe the signal. This helps to weight the posterior distribution based on prior knowledge. In the denominator, there is the Evidence, which represents the marginalization of the numerator over all of parameter space: $p(d) =\int_{\vec{\Theta}} p(d|\vec{\Theta})p(\vec{\Theta})d\vec{\Theta}$. In gravitational-wave applications, the evidence is intractable to calculate directly.  In Markov Chain Monte Carlo (MCMC), the method that will be used to determine the posterior, the evidence enters only as a multiplicative factor. Therefore, we can neglect its direct computation here. The evidence could be computed in MCMC using Stepping-Stone Sampling \cite{Maturana-Russel2019SteppingStone} or Thermodynamic Integration \cite{Goggans2004ThermoInt, Lartillot2006ThermoInt}. 

The key component of Equation~\ref{eq:bayes}, where actual signal modeling and gravitational-wave specific computations are included, is the Likelihood: $\mathcal{L}=p(d|\vec{\Theta})$. The gravitational-wave Likelihood is determined by matching a template, $h(t)$, generated using parameters $\vec{\Theta}$, to the data stream. In a noiseless situation, if $h(t)$ is the correct signal model, the maximum Likelihood of zero would occur where $\vec{\Theta} = \vec{\Theta}_\text{true}$, giving $h(t) = s(t)$. The Likelihood for a specific template is given by
\begin{align}\label{eq:like}
\begin{split}
	\log{\mathcal{L}} \propto  -\frac{1}{2}& \langle d - h|d-h \rangle  \\ = &-\frac{1}{2} \left(  \langle d | d \rangle +  \langle h | h \rangle - 2  \langle d | h \rangle \right),
\end{split}
\end{align}
where we define $ \langle a | b \rangle$ to be the noise-weighted inner product between time-domain data streams $a(t)$ and $b(t)$. This inner product is general and can include non-stationary noise effects by including a proper noise-covariance matrix. Under the assumption of Gaussianity and stationarity, we can write down the inner product necessary for the Likelihood:
\begin{equation}\label{eq:inner}
	 \langle a | b \rangle = 4 \text{Re} \sum_{i = AET}\int_0^\infty \frac{\tilde{a}^{i}(f)^*\tilde{b}^{i}(f)}{S_n^{i}(f)} df,
\end{equation}
where the sum is over the three time-delay interferometry observables $A,E,T$ (see below in Section~\ref{sec:template_gen}). Notice the frequency range is over positive frequencies because the signals are real valued. In this work, we assume $S^i_n(f)$ are known analytic functions, and, therefore, neglect assessing or including their uncertainty. The noise curves for TDI obervables $A$ and $E$ are identical, but $T$ is different. Their formulae are given in \cite{Marsat:2020rtl}. 

The square root of the template-only term in the Likelihood, $\sqrt{\langle h | h \rangle}$, represents the optimal SNR achievable for a given template. In general, the parameters we build the template with will not be exactly equal to the true parameters (where $\mathcal{L}=0$ in the noiseless case). The extracted SNR of a specific template against the data is given by $\langle d | h \rangle / \sqrt{\langle h | h \rangle}$. The extracted SNR will be an important component of the automated extraction pipeline. 

\subsection{Template generation}\label{sec:template_gen}

The main component necessary for computing the Likelihood is the template. The LDC dataset is created by generating the waveform in the frequency domain and then transforming it to the time domain. For this dataset, the \texttt{PhenomD} waveform model \cite{Husa2016, Khan2016}  is used. This model includes only the $l=m=2$ harmonic with aligned spins. In reality, multiple harmonics for MBHBs will be observable. To get a sense of the posterior that would be produced with the inclusion of the higher harmonic modes, we will use the \texttt{PhenomHM} waveform \cite{London2018} by injecting a signal with parameters equivalent to the LDC injection parameters. This will be discussed further in Section~\ref{sec:segment3}.  

The time-domain transform of the LDC injection waveform is then put through the time-domain LISA response function that assumes equal-armlength orbits for the LISA spacecraft. This step projects the signal on to the arms of the LISA constellation and then combines these projections using time-delay interferometry (TDI) to obtain the three TDI observables: $X,\ Y,\ Z$ \cite{Tinto1999, Armstrong1999,Estabrook2000, Dhurandhar2002, Tinto2005}. The LDC provides these three observables. These observables are correlated in their noise properties, so they are transformed into uncorrelated observables: $A,\ E,\ \text{and}\ T$. This transform is given by \cite{Vallisneri2005}
\begin{align}
    A =& \frac{1}{\sqrt{2}}\left(Z-X\right), \\
    E =& \frac{1}{\sqrt{6}}\left(X-2Y+Z\right), \\
    T =&\frac{1}{\sqrt{3}}\left(X+Y+Z\right).
\end{align}
When generating the template, it is these uncorrelated observables that are determined for each Likelihood computation.

For our current pipeline, we will work in the frequency domain and generate the Fourier transform of $h(t)$: $\tilde{h}(f)$. Note that, prior to the analysis, we transform the data to get $\tilde{d}(f)$.  Quasi-circular aligned-spin waveforms, like \texttt{PhenomD} and \texttt{PhenomHM}, require 11 parameters: $\vec{\Theta}= \{M_T,\ q,\ a_1,\ a_2,\ D_L,\ \phi_\text{ref},\  \iota,\ \lambda,\ \beta,\ \psi,\ t_\text{ref}\}$. $M_T$ is the binary total mass: $M_T=m_1 + m_2$. The mass ratio is $q=m_2/m_1$ with $m_2<m_1$. The dimensionless spins, with range $-0.99\leq a_i \leq 0.99$, for $m_1$ and $m_2$ are $a_1$ and $a_2$, respectively. The spins are aligned with the orbital angular momentum of the system: a negative spin indicates anti-parallel alignment. The luminosity distance is given as $D_L$. The reference phase and polarization angles are $\phi_\text{ref}$ and $\psi$, respectively. The orbital inclination is $\iota$. The ecliptic longitude is $\lambda$ and $\beta$ is the ecliptic latitude. The time of coalescence is $t_\text{ref}$. In this work, $t_\text{ref}$ and $\phi_\text{ref}$ are both set at $f_\text{ref}=f_\text{max}$, where $f_\text{max}$ is a value determined internally in the \texttt{PhenomD} code where the value of $f^2A_{22}(f)$ is maximized ($A_{22}$ is the amplitude of the (2,2) mode). All of the extrinsic parameters given here are defined with respect to the LISA constellation reference frame.

With the intrinsic parameters and the luminosity distance, we can generate the source frame waveform scaled for the distance as the amplitude, $A(f)$, and phase, $\phi(f)$: 
\begin{equation}\label{eq:amp_and_phase}
    \tilde{h}_{lm}(M_T, q, a_1, a_2, D_L) = A_{lm}(f)e^{-i\phi_{lm}(f)},
\end{equation}
where $(l,m)$ are the harmonic mode indices.\footnote{Even though the LDC waveform only contains the $l=m=2$ mode, we include the generic description that includes higher harmonics in preparation for the higher harmonic posterior presented in Section~\ref{sec:segment3}.} 

The scaled source-frame waveform must then be transformed into the TDI observables through the frequency-domain transfer function, $\mathcal{T}(f, t_{lm}(f))$. This function is described in detail in \cite{Marsat:2018oam, Marsat:2020rtl}. It is determined by the extrinsic parameters $\{\phi_\text{ref},\  \iota,\ \lambda,\ \beta,\ \psi,\ t_\text{ref}\}$ and is time- and frequency-dependent due to the evolution of the LISA constellation throughout its orbit. The time-frequency dependence for each harmonic, $t_{lm}(f)$, is determined from the stationary-phase approximation (SPA) used to produce the frequency-domain waveforms by taking the derivative of the phase with respect to frequency:
\begin{equation}\label{eq:tf}
    t_{lm}(f) = t_\text{ref} - \frac{1}{2\pi}\frac{d\phi_{lm}(f)}{df}.
\end{equation}
The LISA response depends on these extrinsic parameters given in the Solar System Barycenter (SSB) reference frame. The sampler, discussed in Section~\ref{sec:pipeline}, generates posterior samples in the LISA constellation reference frame. These parameters are converted to the SSB frame prior to the Likelihood computation. This conversion is discussed in \cite{Marsat:2020rtl}. 

By combining $h_{lm}(f)$ and $\mathcal{T}(f, t_{lm}(f))$ within each harmonic, we produce the templates for each TDI channel:
\begin{equation}\label{eq:response}
    \tilde{h}^{A,E,T}(f) = \sum_{lm} \mathcal{T}^{A,E,T}(f, t_{lm}(f))\tilde{h}_{lm}(f).
\end{equation}

\subsection{Heterdyned Likelihoods}

For two out of the three sections of the pipeline, we take advantage of the Heterodyned Likelihood technique \cite{Cornish:2010kf, Cornish:2021lje}. The specific linear formulation implemented for this work can be found in \cite{Zackay:2018qdy}. We will refer to this fast Likelihood technique as ``Heterodyning'' for the remainder of the paper to avoid confusion with other LISA-related literature, but please note the specific maths and formulation presented below are often referred to as ``Relative Binning,'' the name provided in \cite{Zackay:2018qdy}. We will briefly summarize the Heterodyning technique and its specific implementation used here. We refer the interested reader to \cite{Cornish:2021lje, Zackay:2018qdy} for more information. 

Heterodyning simplifies the frequency-domain Likelihood computation by operating on residuals between templates, rather than direct comparisons of templates to data. This involves precomputing a handful of quantities representing the contribution to the Likelihood from $\langle d|h\rangle$ and $\langle h | h \rangle$ terms at the full frequency resolution of the Fourier transform of the data stream.

We assume that a ``reference template,'' $\tilde{h}_0(f)$, has been found that is similar in its characteristic shape and properties to the true signal. This reference template is compared to the data at the Fourier frequencies. Online computations will involve computing new templates, $\tilde{h}(f)$, that are close in the waveform domain to the reference template. Discretizing the inner product and using the residual ratio between the two templates, $\tilde{r}(f) = \tilde{h}(f)/\tilde{h}_0(f)$, gives 
\begin{align}\label{eq:discretized}
\begin{split}
	\langle d |  h \rangle \approx Z\left[ \tilde{d}(f), \tilde{h}(f) \right]&  \\ = 4\text{Re}&\sum_f \frac{\tilde{d}(f)\tilde{h}_0^*(f)}{S_n(f)} \tilde{r}^*(f)\Delta f.
\end{split}
\end{align}
$Z$ here represents the heterodyned approximation to the inner product value. A similar equation is given for the $\langle h | h\rangle$ term in the Likelihood. In practice, $\tilde{r}(f)$ can be smoothly approximated over a large (relative to the Fourier bin width) frequency bin, $b$,  as a complex linear segment: 
\begin{equation}
	\tilde{r}(f) = \frac{\tilde{h}(f)}{\tilde{h}_0(f)}= \tilde{r}_0(f) + \tilde{r}_1(f) (f - f_b) + \mathcal{O}(f^2),
\end{equation}
where $f_b$ is the frequency at the center of bin $b$. Equation~\ref{eq:discretized} can then be recast over large frequency bins while including $\tilde{r}(f)$ up to linear order:
\begin{align}\label{eq:into_bins}
	Z\left[\tilde{d}(f), \tilde{h}(f) \right] = 4\text{Re}\sum_b \sum_{f \in b} \frac{\tilde{d}(f)\tilde{h}_0^*(f)}{S_n(f)}&  \\  \times(\tilde{r}_0^*(f) + \tilde{r}_1^*(f)(f - f_b))&\Delta f.
\end{align}
Since the terms in $\tilde{r}(f)$ are constant over the bin, they can be separated from the within-bin summation:
\begin{align}\label{eq:d_h_rel_bin}
\begin{split}
	Z\left[\tilde{d}(f), \tilde{h}(f) \right] &= \\ 4\text{Re}&\sum_b A_0(b)\tilde{r}_{0,b}^* + A_1(b)\tilde{r}_{1,b}^*\ ,\quad \text{with}
\end{split}\\
	A_0(b) &= \sum_{f\in b}\frac{\tilde{d}(f)\tilde{h}_0^*(f)}{S_n(f)} \Delta f \qquad \text{and} \\
	A_1(b) &= \sum_{f\in b}\frac{\tilde{d}(f)\tilde{h}_0^*(f)}{S_n(f)}(f-f_b) \Delta f.
\end{align}
The complimentary equation for $\langle h | h\rangle$ is given by
\begin{align}\label{eq:h_h_rel_bin}
\begin{split}
	Z\left[\tilde{h}(f), \tilde{h}(f) \right] &= \\ 4\text{Re}\sum_b \Big\{\Big.&B_0(b)\left|\tilde{r}_{0,b}\right|^2 \\ &+ 2B_1(b)\left(\text{Re}\left[\tilde{r}_{0,b}\tilde{r}_{1,b}^*\right]\right)\Big.\Big\}\ ,\quad \text{with}
\end{split} \\
	B_0(b) &= \sum_{f\in b}\frac{\left|\tilde{h}_0(f)\right|^2}{S_n(f)} \Delta f \qquad \text{and} \\
	B_1(b) &= \sum_{f\in b}\frac{\left|\tilde{h}_0(f)\right|^2}{S_n(f)}(f-f_b) \Delta f.
\end{align}

The tangible reason for using Heterodyning is the strong reduction in the number of individual frequencies required for an accurate Likelihood computation. Typical LISA data streams are of order $\sim10^6-10^7$ points. Heterodyned Likelihoods use $\sim$100s of points. During online computations, $\tilde{h}(f)$ is determined at the edges of each large frequency bin. The terms in $\tilde{r}(f)$ to linear order are then computed by dividing the new waveform by the reference waveform, and then determining the slope and intercept of the complex line stretching across each frequency bin. 

The main difference between the Likelihood computation in this work compared to \cite{Zackay:2018qdy} is the inclusion of the LISA response. The response function is sufficiently smooth that it does not harm the ``closeness'' assumption that the residual waveform can be smoothly represented as a complex linear function over a larger frequency bin.

 \subsection{A note on the use of Graphics Processing Units}
 
 All codes used in this work for Likelihood computations, including Heterodyning, waveform generation, and the LISA response function, are performed on Graphics Processing Units (GPU). GPUs strongly accelerate these computations. A general description for the GPU coding architecture can be found in \cite{Katz:2020hku}. There have been some crucial improvements from the codes described in \cite{Katz:2020hku}. The main difference is that the waveform and response codes, which were formerly run in parallel on CPU cores, are now redesigned for large batches of computations on the GPUs. This is not a \textit{major} change as the core of the code remains constant. Rather, the outer layers of the waveform and response codes were adjusted to facilitate proper usage of GPU parallelization across GPU blocks and threads. This crucial change creates the immense speed of the batched Heterodyning technique. The timing of the various Likelihood computations will be discussed in the next section. The MBHB waveform and Likelihood code is publicly available here: \href{https://github.com/mikekatz04/BBHx}{github.com/mikekatz04/BBHx} \cite{michael_katz_2021_5737837}. All other codes used in this work are available upon request to the author. They will eventually be made public.
 
\section{Automated extraction pipeline} \label{sec:pipeline}

The pipeline consists of three distinct modules: an initial search phase starting across the prior volume; a refining search phase to locate the maximum Likelihood point and build up the high Likelihood population; and a parameter estimation portion to produce the posterior. Each module passes information to the next flowing in a fully-automated way from module to module. 

All modules use the same priors. They are given in Table~\ref{tb:priorinfo}. Uniform distributions are used on parameters $\{q, a_1, a_2, D_L, \phi_\text{ref}, \lambda, \psi, t_\text{ref}\}$. The total mass prior is log-uniform. The $\iota$ prior is uniform in $\cos{\iota}$, while the prior on $\beta$ is uniform in $\sin{\beta}$. 

\begin{center}
\begin{table}[h!]
\begin{tabular}{| c | c | c |}
 \hline
Parameter & Lower Bound & Upper Bound \\
\hline
$\ln{M_T}$ & $\ln{\left(10^5\right)}$& $\ln{\left(10^8\right)}$ \\
\hline
$q$ & $5\times10^{-2}$ & 1 \\
\hline
$a_1$ & -1 & 1 \\
\hline
$a_2$ & -1 & 1 \\
\hline
$D_L$ & $10^{-2}$ & $10^{3}$ \\
\hline
$\phi_\text{ref}$ & 0 & $2\pi$ \\
\hline
$\cos{\iota}$ & -1 & 1 \\
\hline
$\lambda$ & 0 & $2\pi$ \\
\hline
$\sin{\beta}$ & -1 & 1 \\
\hline
$\psi$ & 0 & $\pi$ \\
\hline
$t_\text{ref}$ & $0$ & $T_\text{obs}$ \\
\hline
\end{tabular}
\caption{The lower and upper bounds of the prior distributions used in all modules of the pipeline presented in this work. All priors are uniform distributions, except for the prior on $M_T$, $\iota$, and $\beta$. The prior for $M_T$ is log-uniform. The prior on $\iota$ and $\beta$ are uniform distributions on $\cos{\iota}$ and $\sin{\beta}$. The sky and orientation parameters are given in the LISA reference frame. $T_\text{obs}$ is the observation time. For the LDC datasets, $T_\text{obs}\sim1.33$ years.}\label{tb:priorinfo}
\end{table}
\end{center}

All pipeline modules also use the same MCMC sampler. This sampler is based on the overall structure of \texttt{emcee} \cite{emcee} with the parallel tempering scheme from \texttt{ptemcee} \cite{Vousden2016}. The specific implementation used here for the parallel tempering method is given in \cite{Katz:2020hku}. There are two proposals used. The first is the typical ``stretch'' proposal described in \cite{Goodman2010}, which is used for $\sim$85\% of proposals. The second is a sky-mode hopping proposal. As explained in \cite{Marsat:2020rtl}, when sampling in the LISA reference frame, as is done in this work, there exists 8 distinct sky modes in terms of the extrinsic parameters. There are four longitudinal modes at $\{\lambda + (0, 1,2,3) \pi/ 2, \psi + (0, 1,2,3) \pi/ 2\}$ and two latitudinal modes: $\{\pm\beta, \pm\cos{\iota}, \pm\cos{\psi})\}$. Three varieties of this proposal are used. The first is to move the walker to a mode drawn randomly with replacement from all eight sky modes. This is used for $\sim$4\% of proposals. Proposals to change just the longitudinal or latitudinal mode are used $\sim$3\% and $\sim$8\% of the time, respectively. The relative percentages of the sky-mode hopping proposals reflect that at higher frequencies, the eight modes will reduce to just the two latitudinal modes at the proper ecliptic longitude. In this situation, longitudinal mode-hopping proposals are not efficient.

\subsection{Module 1: Direct full-template search}\label{sec:segment1}

Harnessing the power of GPUs allows for a fast and direct Likelihood computation against the data as described in \cite{Katz:2020hku}. This computation takes $\sim$few milliseconds. We use 80 walkers per temperature with 4 temperatures ranging from one (target distribution) to infinity (prior). Only 4 temperatures are used (compared to 10 for the later segments) due to the longer Likelihood computation time as compared to the Heterodyning methods used below. This setup allows the chains to take more steps in the same amount of wall time, which was found to be advantageous for the initial search. The starting positions of all walkers are generated randomly from the prior distribution. These positions are then evolved with the MCMC techniques given above. This procedure is used to obtain samples from the posterior, which will be preferentially located towards higher Likelihood values. The goal of this portion of the pipeline is to locate the proper frequency band where the signal lies. In other words, this means roughly locating the true value on the total mass. This is crucial to the success of the Heterodyned Likelihood since it can give erroneous results at frequencies that are not mutually included in both the reference template ($h_0$) and tested template ($h$). 

\begin{figure}[t]
\begin{center}
\includegraphics[scale=0.45]{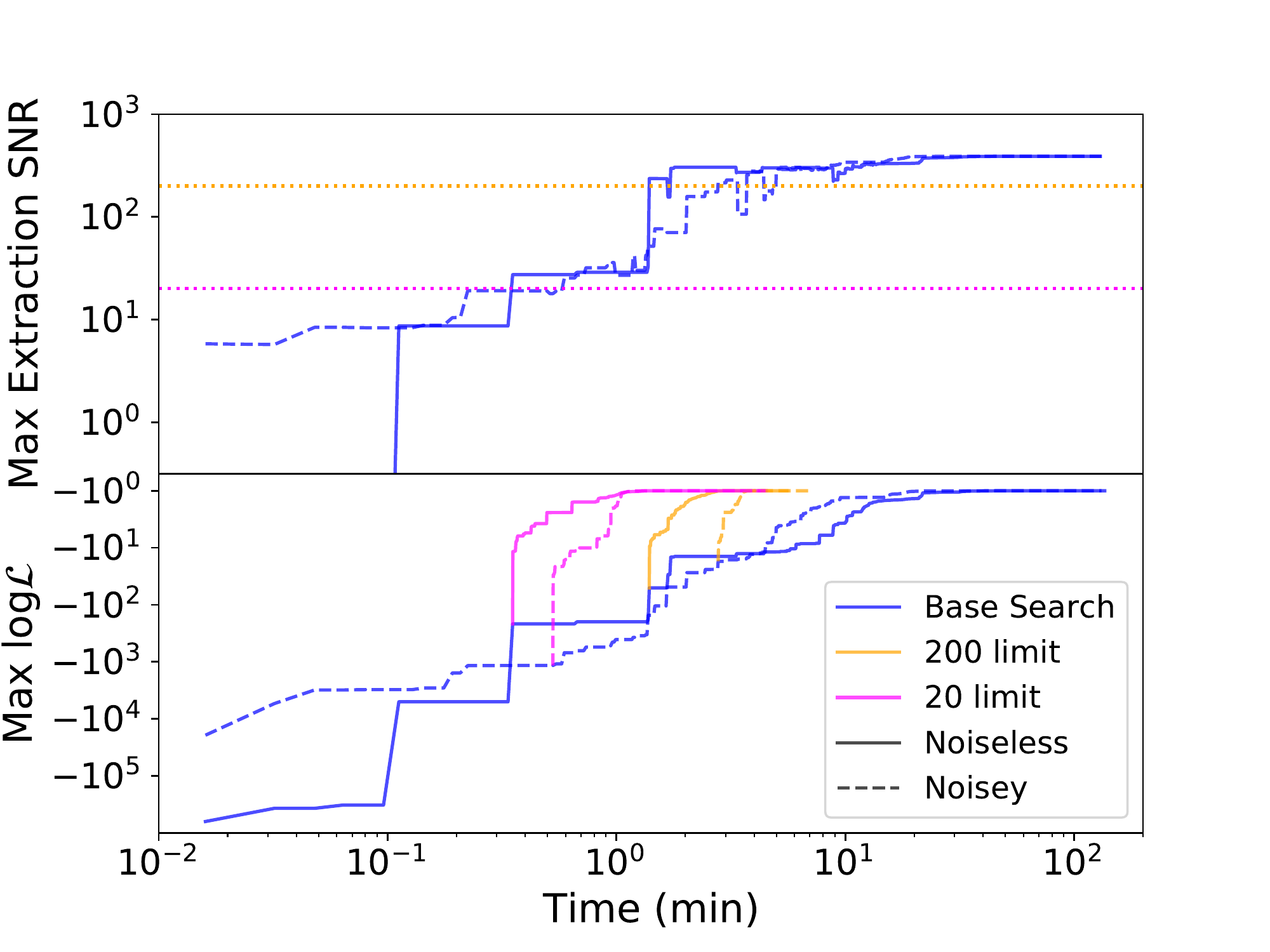}
\caption{The evolution of the maximum log-Likelihood and extraction SNR over time as the base search module proceeds. The base search runs for the noiseless and noisy cases are shown in blue solid and dashed lines, respectively. The noisy log-Likelihood is normalized so that the maximum log-Likelihood found by the sampler is set to zero, equivalent to the noiseless maximum log-Likelihood. The base search module culminates when a user-specified extraction SNR is found by the sampler. Two extraction SNR limits are shown here: 20 (magenta) and 200 (orange). In the top panel, these limits are marked with dotted lines to indicate the turnover point: when the blue line reaches these horizontal lines for the first time, it will end sampling in the base search module and move on to the second module. In the bottom panel, the evolution of the log-Likelihood during the second module is also shown over time for both extraction SNR limits. These lines begin at the moment when the initial search reaches the desired extraction SNR and turns over into the second module; they end at the time when the pipeline has converged on a maximum Likelihood point.}\label{fig:search_info}
\end{center}
\end{figure}

This portion of the pipeline takes $\sim$minutes or less to complete. The sampler is outfit with a ``stopping'' function that performs a test every 150 iterations. The test is based on the extracted SNR (see Section~\ref{sec:bayes_mbhb}). When it reaches a user-specified value, the sampling ends and passes the data file information to the next pipeline module. The evolution of the sampler in this module is shown in Figure~\ref{fig:search_info} in terms of the Likelihood and the extracted SNR over time. In both the noisy and noiseless cases, the sampler climbs the Likelihood surface quickly and consistently. This figure also includes a rough visualization of the different durations of this module for two different extraction SNR convergence limits (20 and 200). In practice, a lower extracted SNR threshold ($\sim10-20$) will be used to terminate the first module in the pipeline given that the SNR of the signal in real data will be unknown. However, a higher threshold could be used to further refine parameters with this direct method if desired, as long as the empirically determined extracted SNR continues to rise. This may be desirable if the Heterodyned Likelihood has issues with the determined frequency band over which the calculation is performed. Further refining the parameters may give a more accurate frequency band depending on the specific source. In the case examined here, the lower extracted SNR threshold (20) is successful in finding a proper frequency band over which the Heterodyning computation is performed.

An advantage of this method is that it is completely general (given noise and detector assumptions) to frequency domain waveforms. Therefore, it can be easily extended to include precession, eccentricity, etc. This module could be run all the way through posterior production if desired, but this is not as efficient (by a factor of $\sim20-100$) as the chosen fast Likelihood methods described in the following sections. 

\subsection{Module 2: Heterodyned search}\label{sec:segment2}

After reaching the extracted SNR threshold, the highest Likelihood point is used as a reference template for Heterodyning. The frequency content contained in the reference template over the observation time, $T_\text{obs}$, is set as the frequency band over which the Heterodyning method is applied ($\tilde{h}_0(f)\neq0$). The frequency grid is chosen to have 128 log-spaced frequencies. With these settings, each Likelihood computation takes $\sim5\mu$s. The reference template is updated every 5000 iterations of the sampler by replacing it with the highest Likelihood point in the population. 

This segment of the pipeline uses 400 walkers in 10 temperatures (4000 total), taking advantage of the immense power of the low-memory, batched Heterodyned Likelihood computations.

Each walker is assigned an initial position from the top 4000 Likelihood points resulting from Module 1 and is evolved forward according to the same MCMC method as Module 1. The stopping function for this section of the pipeline is determined by the convergence of the maximum Likelihood located by the sampler. Every 150 iterations of the sampler, the stopping function checks the maximum Likelihood point. If this point does not change its value by 1 during multiple consecutive checks, the sampling is stopped and the data is saved and passed on to the parameter estimation portion in the final module. The number of consecutive checks necessary to stop is a user-defined parameter that is set in this work to 30. This is conservative allowing for $30\times150$ iterations $=4.5\times10^3$ iterations of the sampler after finding the initial convergent Likelihood value. Typically, this segment is $\sim$minutes, depending on the exact stopping criterion settings. The evolution of the log-Likelihood for Module 2 with different extraction SNR limits from Module 1 is shown in Figure~\ref{fig:search_info} as the magenta (20 limit) and orange (200 limit) lines in the bottom panel. As the tracks flatten out near $\log{\mathcal{L}}\approx0$, the maximum Likelihood begins to converge. The length of this flat portion gives the amount of time required for the convergence criterion to be met ($\sim$minutes). Ultimately, the use of Heterodyning in Module 2 improves the speed of the search by a factor of $\sim20-100$. The higher end of that range indicates how long it takes the base search in Module 1 to satisfy the log-Likelihood convergence criterion of Module 2. 

\subsection{Module 3: Heterodyned parameter estimation}\label{sec:segment3}

The final module generates the posterior starting from information around the maximum Likelihood point obtained during Module 2. The exact same procedure is used from Module 2 in terms of the Heterodyned Likelihood settings, timing, and updates. A burn-in of 1000 iterations is imposed. The sampler then collects posterior samples for $5\times10^4$ iterations. With 4000 walkers, this produces $4\times 10^8$ posterior evaluations with $4\times10^7$ evaluations along the cold chain probing the target distribution.

\begin{figure*}[tbh]
\begin{center}
\includegraphics[scale=0.5]{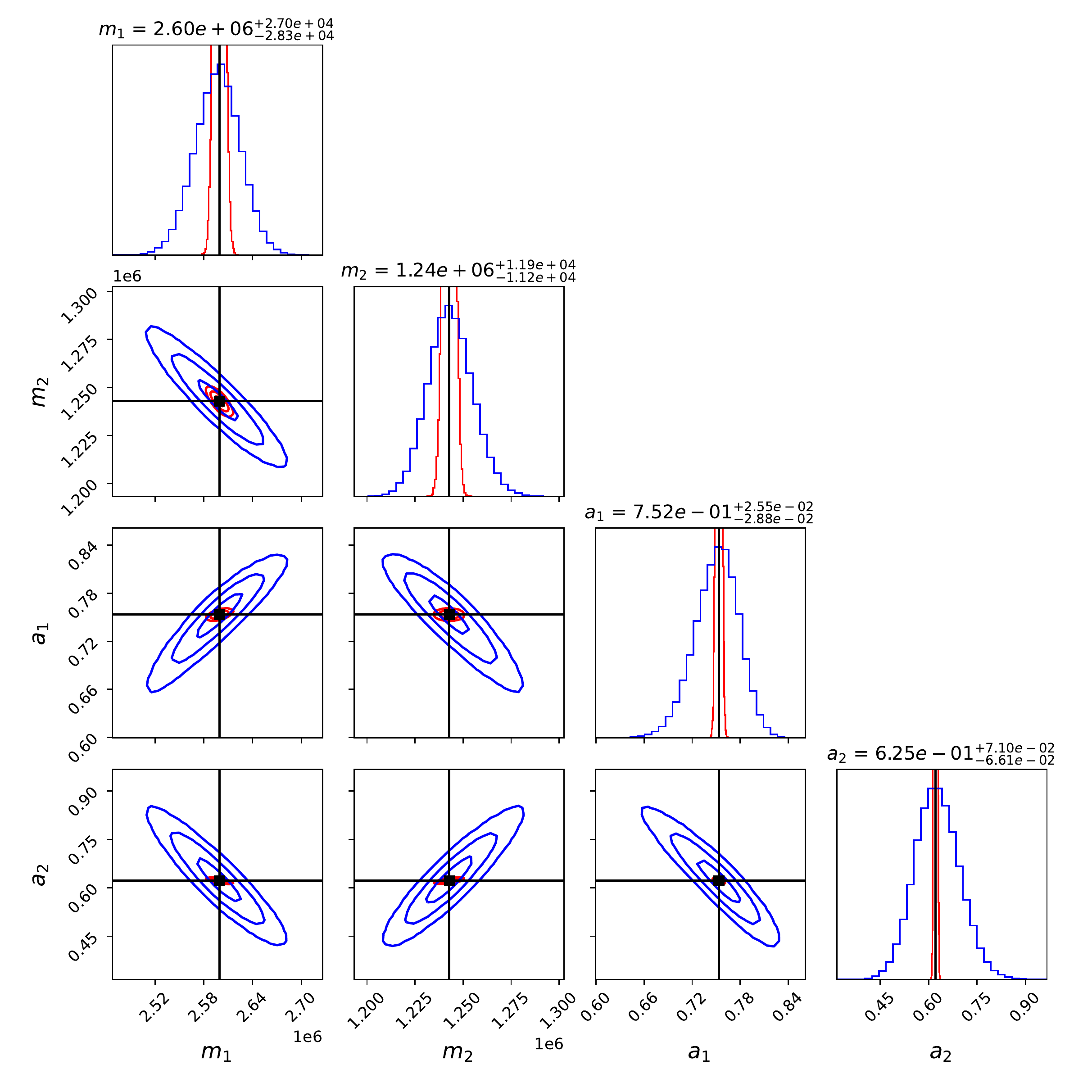}
\caption{The posterior distribution on the intrinsic parameters of the source in the noiseless LDC dataset. The blue contours represent the main posterior distribution. The red distribution is shown to give an idea of more realistic posteriors that include higher-order harmonic modes. The true injection values are marked in black. The posterior distributions contain the $1\sigma$, $2\sigma$, and $3\sigma$ contours.}\label{fig:noiseless_intrinsic}
\end{center}
\end{figure*}

\begin{figure*}[tbh]
\begin{center}
\includegraphics[scale=0.5]{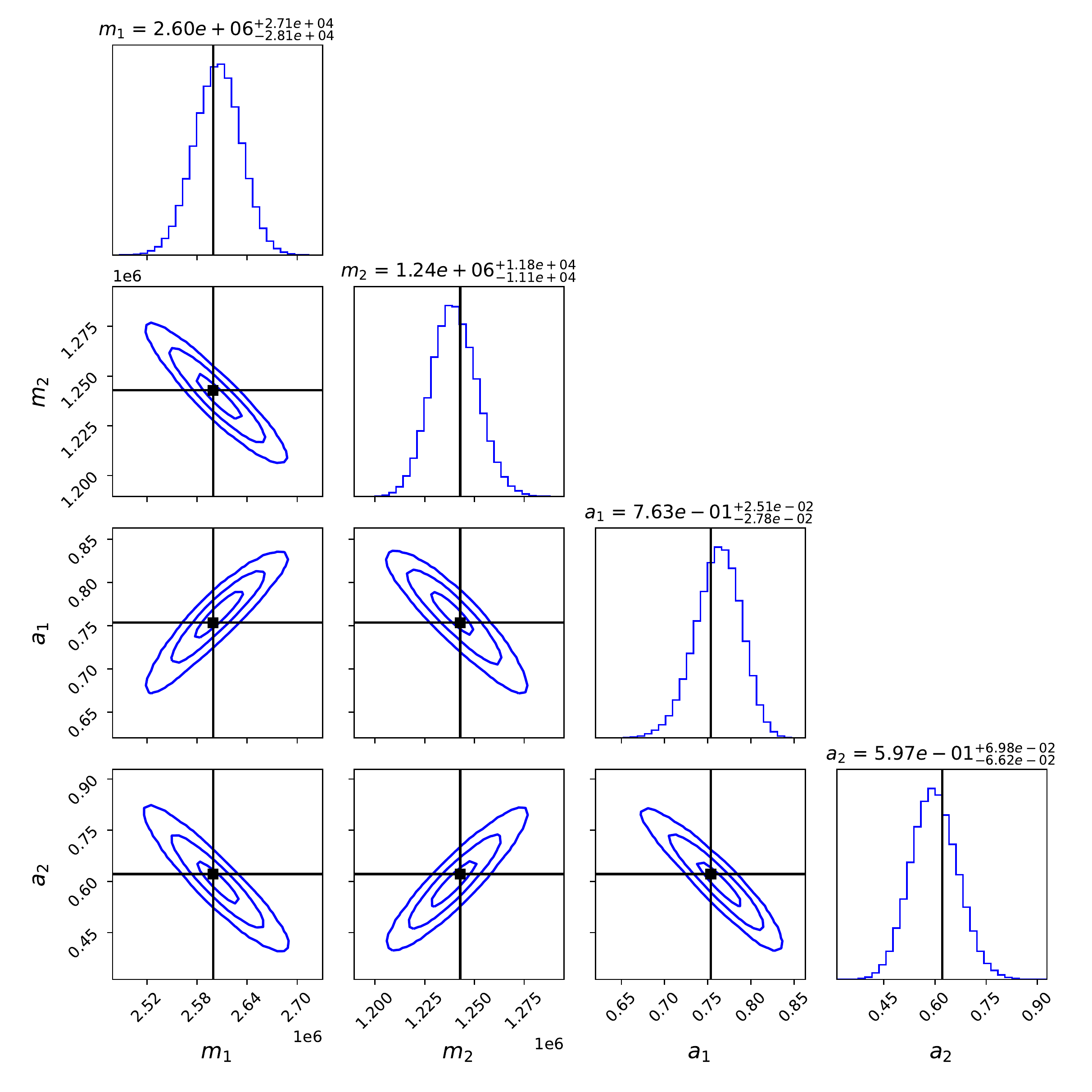}
\caption{The posterior distribution on the intrinsic parameters from the noisy LDC dataset represented with blue contours. The true injection values are marked in black. The posterior distribution contains the $1\sigma$, $2\sigma$, and $3\sigma$ contours.}\label{fig:noise_intrinsic}
\end{center}
\end{figure*}

\begin{figure*}[tbh]
\begin{center}
\includegraphics[scale=0.55]{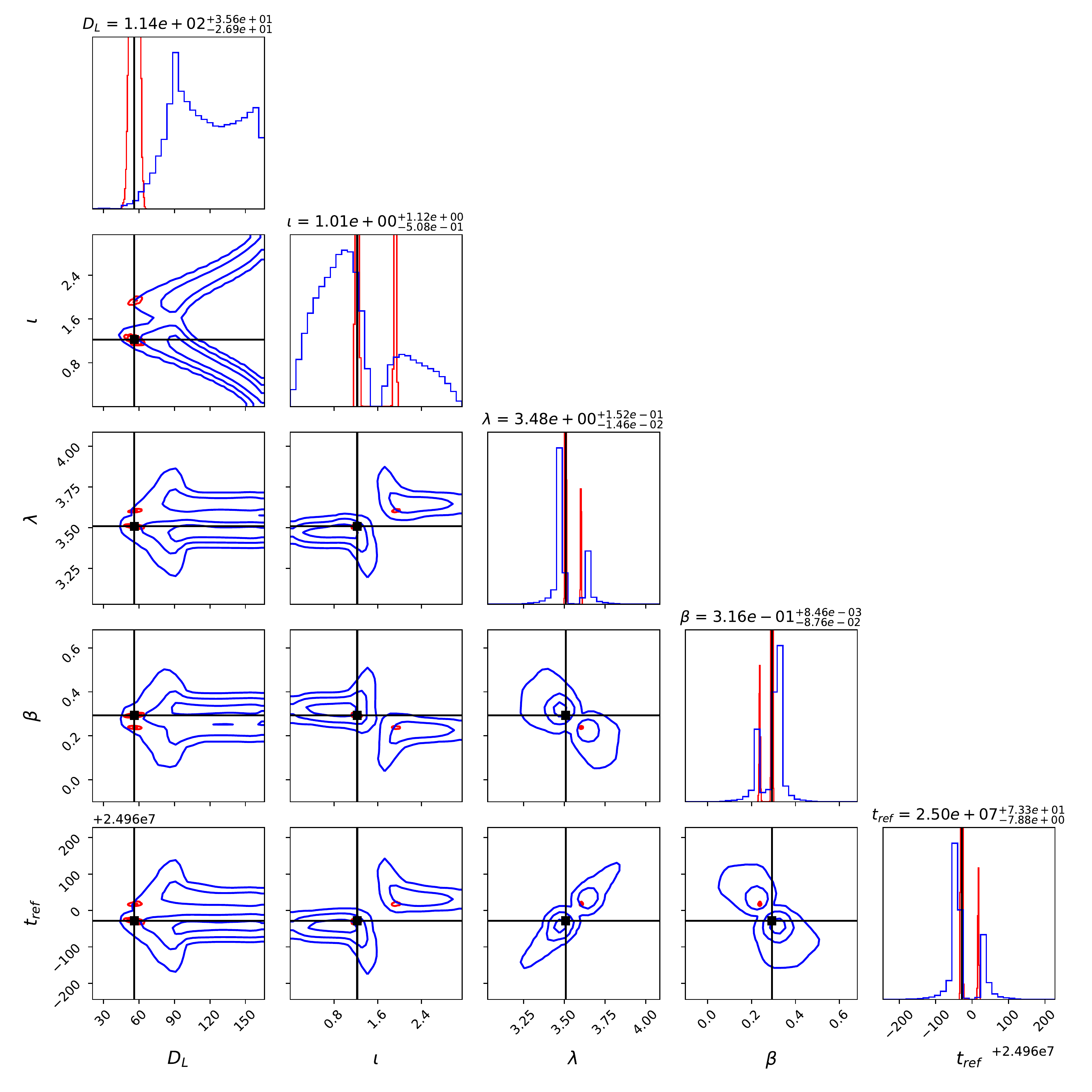}
\caption{The posterior distribution, shown as blue contours, on the extrinsic parameters of the source in the noiseless LDC dataset. The sky and orientation parameters given here are in the SSB frame. The red distribution represents the posterior from the \texttt{PhenomHM} injection and template fitting. The true injection values are marked in black. The posterior distributions contain the $1\sigma$, $2\sigma$, and $3\sigma$ contours.}\label{fig:noiseless_extrinsic}
\end{center}
\end{figure*}

\begin{figure*}[tbh]
\begin{center}
\includegraphics[scale=0.55]{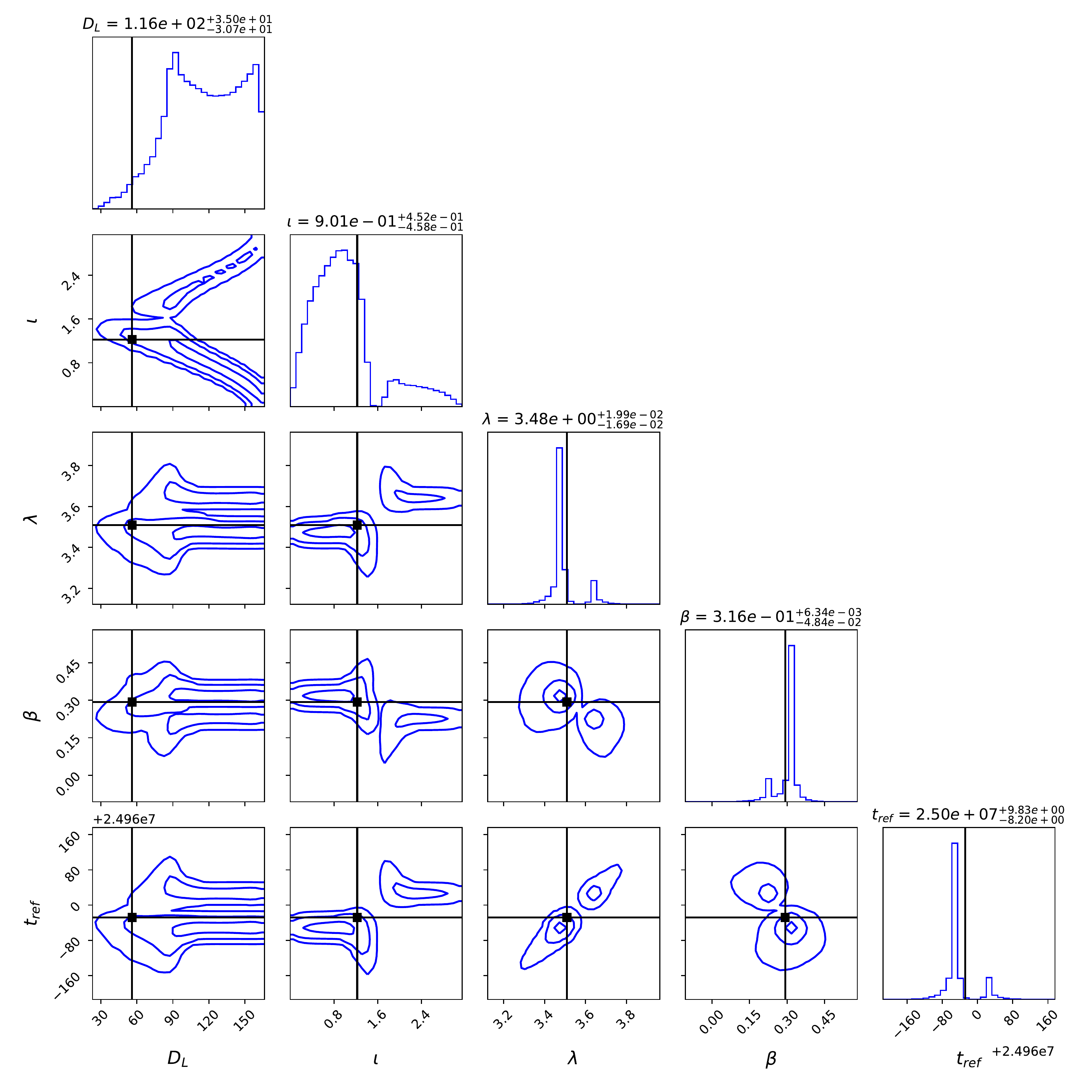}
\caption{The posterior distribution on the extrinsic parameters from the noisy LDC dataset represented with blue contours. The sky and orientation parameters are shown in the SSB reference frame. The true injection values are marked in black. The posterior distribution contains the $1\sigma$, $2\sigma$, and $3\sigma$ contours.}\label{fig:noise_extrinsic}
\end{center}
\end{figure*}

The posterior samples for the noise-infused and noiseless cases are shown in Figures~\ref{fig:noiseless_intrinsic} and \ref{fig:noise_intrinsic}, respectively, for the intrinsic parameters. Figures~\ref{fig:noiseless_extrinsic} and \ref{fig:noise_extrinsic} display the posterior distribution on the extrinsic parameters for the noiseless and noisy cases, respectively. The mean and standard deviation for unimodal parameters in all tests are given in Table~\ref{tb:means}. The parameters of the maximum Likelihood point attained from the sampler during each MCMC run are shown in Table~\ref{tb:maxll}. The posterior information for the waveform with higher-order modes is also added to the tables and figures. Extrinsic parameters in these figures and tables are given in the SSB frame.

The injected signal has a strong SNR of $\sim390$. Therefore, the constraints on intrinsic parameters are quite strong, even for the case where only the (2,2) mode is included. The sky location is fairly constrained and exhibits the character expected. This signal reaches high enough frequencies that the base eight-fold sky degeneracy in the LISA reference frame is reduced to just the latitudinal modes at the true longitude. The two-dimensional sky posterior shows the transformation of this characteristic feature to the SSB frame. The two sky modes also lead to multimodal distributions in the inclination and reference time (in the SSB frame). The most likely value for $t_\text{ref}$ found by the sampler is approximately 30 seconds before the value given in the LDC dataset. This is due to a differing convention in the reference frequency at which $t_\text{ref}$ is assigned. In the LDC, $t_\text{ref}$ is set at the highest frequency in the waveform. As described in Section~\ref{sec:template_gen}, we set $t_\text{ref}$ at $f_\text{max}$, where $f^2A_{22}(f)$ reaches a maximum. The difference in $t_\text{ref}$ between these two conventions for this specific source is $\sim30$ seconds, which confirms we have recovered the correct value. The distance posterior in both the noise and noiseless cases are typical for sources with the (2,2)-mode-only waveforms. The true value is closer to the observer than the mean and mode of the distribution. 

The effect of the noise is also visible. This particular noise realization produces only a small shift in the noiseless distribution. The parameter posterior from the noisy data is qualitatively and quantitatively similar to the noiseless distribution in all parameters.

It is clear, from examining the distributions over time as more samples are collected, that intrinsic parameters converge rapidly while the sky and orientation parameters converge more slowly. A posterior distribution in the extrinsic parameters roughly equivalent to the final distribution is found as fast as the intrinsic parameters are located; however, it does take significantly more time for the sampler to fully converge the tails in the extrinsic parameter distributions.

In order to better inform the reader of expected and more realistic parameter constraints and posterior distributions, we provide a posterior that shows the effect of analyzing higher-order harmonic modes for the same source that is used for the LDC-1A dataset. To obtain the higher-mode distributions, we injected a \texttt{PhenomHM} signal with the same parameters as the LDC source and then used the brute-force Likelihood approach from Section~\ref{sec:segment1} subbing in \texttt{PhenomHM} for \texttt{PhenomD}. The sampling begins with spreading the walkers around the true injection point. 

The higher-mode distribution is better in constraining parameters by about an order of magnitude. The higher harmonics also strongly inform the marginalized distance posterior to the point where it is now Gaussian and spanning the true value as its mean. This is due to the broken degeneracy between inclination and distance that exists in the (2,2)-only case. This broken degeneracy also improves sampling efficiency. See \cite{Marsat:2018oam} for more detail on comparing sampling effects and posteriors from (2,2)-mode-only waveforms versus waveforms with higher modes. 

\begin{center}
\begin{table*}
\begin{tabular}{| C{2.0cm} | L{2.5cm} | L{3.2cm} | L{3.2cm} | L{3.2cm} |} 
 \hline
  Parameter & \multicolumn{1}{c|}{Injection}  & \multicolumn{1}{c|}{(2,2) Noiseless} & \multicolumn{1}{c|}{(2,2) Noise} & \multicolumn{1}{c|}{HM Noiseless} \\ 
 \hline
 \hline
$m_1$ & $2.599137\times10^{6}$ & $2.60\times10^{6} \substack{+3\times10^{4} \\ -3\times10^{4}}$ & $2.60\times10^{6} \substack{+3\times10^{4} \\ -3\times10^{4}}$ & $2.599\times10^{6} \substack{+5\times10^{3} \\ -5\times10^{3}}$ \\ [1.0ex] 
\hline
$m_2$ & $1.242861\times10^{6}$ & $1.24\times10^{6} \substack{+1\times10^{4} \\ -1\times10^{4}}$ & $1.24\times10^{6} \substack{+1\times10^{4} \\ -1\times10^{4}}$ & $1.243\times10^{6} \substack{+3\times10^{3} \\ -3\times10^{3}}$ \\ [1.0ex] 
\hline
$a_1$ & $7.534822\times10^{-1}$ & $7.6\times10^{-1} \substack{+3\times10^{-2} \\ -2\times10^{-2}}$ & $7.5\times10^{-1} \substack{+3\times10^{-2} \\ -2\times10^{-2}}$ & $7.53\times10^{-1} \substack{+3\times10^{-3} \\ -2\times10^{-3}}$ \\ [1.0ex] 
\hline
$a_2$ & $6.215875\times10^{-1}$ & $6.0\times10^{-1} \substack{+7\times10^{-2} \\ -7\times10^{-2}}$ & $6.2\times10^{-1} \substack{+7\times10^{-2} \\ -7\times10^{-2}}$ & $6.22\times10^{-1} \substack{+3\times10^{-3} \\ -3\times10^{-3}}$ \\ [1.0ex] 
\hline
$D_L$ & $5.600578\times10^{1}$ & $1.1\times10^{2} \substack{+3\times10^{1} \\ -4\times10^{1}}$ & $8.76\times10^{1} \substack{+3\times10^{1} \\ -3\times10^{1}}$ & $5.6\times10^{1} \substack{+3\times10^{0} \\ -3\times10^{0}}$ \\ [1.0ex] 
\hline

\end{tabular}
\caption{The mean and standard deviation values associated with the various tests performed. These values are only shown for the mass, spin, and distance parameters because these marginalized distributions are unimodal. The sky and orientation parameters are all multimodal indicating the observed mean value is not representative of the true source. From left to right, the values stated apply to the LDC dataset without noise, the LDC dataset with noise, and the self-injected, noiseless higher modes dataset analyzed with \texttt{PhenomHM}.}\label{tb:means}
\end{table*}
\end{center}

\begin{center}
\begin{table*}
\begin{tabular}{| C{2.0cm} | L{2.5cm} | L{2.5cm} | L{2.5cm} | L{2.5cm} |} 
 \hline
  Parameter & \multicolumn{1}{c|}{Injection}  & \multicolumn{1}{c|}{(2,2) Noiseless} & \multicolumn{1}{c|}{(2,2) Noise} & \multicolumn{1}{c|}{HM Noiseless} \\ 
 \hline
 \hline
$m_1$ & $2.599137\times10^{6}$ & $2.615348\times10^{6}$ & $2.597597\times10^{6}$ & $2.599091\times10^{6}$ \\ [1.0ex] 
\hline
$m_2$ & $1.242861\times10^{6}$ & $1.235689\times10^{6}$ & $1.243229\times10^{6}$ & $1.243054\times10^{6}$ \\ [1.0ex] 
\hline
$a_1$ & $7.534822\times10^{-1}$ & $7.721731\times10^{-1}$ & $7.512426\times10^{-1}$ & $7.539068\times10^{-1}$ \\ [1.0ex] 
\hline
$a_2$ & $6.215875\times10^{-1}$ & $5.716351\times10^{-1}$ & $6.256857\times10^{-1}$ & $6.213842\times10^{-1}$ \\ [1.0ex] 
\hline
$D_L$ & $5.600578\times10^{1}$ & $2.859041\times10^{1}$ & $6.923802\times10^{1}$ & $5.620339\times10^{1}$ \\ [1.0ex] 
\hline
$\iota$ & $1.224532\times10^{0}$ & $1.403617\times10^{0}$ & $1.169769\times10^{0}$ & $1.223004\times10^{0}$ \\ [1.0ex] 
\hline
$\lambda$ & $3.509097\times10^{0}$ & $3.519460\times10^{0}$ & $3.487416\times10^{0}$ & $3.509118\times10^{0}$ \\ [1.0ex] 
\hline
$\beta$ & $2.927012\times10^{-1}$ & $2.655787\times10^{-1}$ & $2.870121\times10^{-1}$ & $2.931597\times10^{-1}$ \\ [1.0ex] 
\hline

$t_{ref}$ & $2.495997\times10^{7}$ & $2.495996\times10^{7}$ & $2.495997\times10^{7}$ & $2.495997\times10^{7}$ \\ [1.0ex] 
\hline

\hline
\end{tabular}
\caption{Parameters at the maximum Likelihood values found by the sampler for the three different tests are shown. These values are given for the LDC dataset without noise, the LDC dataset with noise, and the self-injected higher modes dataset, from left to right respectively. Both runs without noise located a maximum log-Likelihood $\geq-0.2$. If the $\langle d | d \rangle$ term is removed from the Likelihood (i.e. $\langle d | d \rangle$=0), then the maximum log-Likelihood located by the sampler for the run containing noise was $\sim+76005$.}\label{tb:maxll}
\end{table*}
\end{center}

\section{Discussion}\label{sec:discuss}

This pipeline is designed for generic, quasi-circular frequency domain waveforms. It will not require changes to handle precessing and/or eccentric binaries because it does not rely on any type of maximization over parameters or F-statistic-like computations. However, these types of maximization techniques could be added for further efficiency during the initial base search. This type of change should not affect the Heterodyned Likelihood sections of the pipeline because the maximization techniques will maintain a strong determination of the total mass. This is essential to the success of the Heterodyned Likelihood modules because Heterodyning requires an accurate estimation of the frequency band containing the waveform. 

While the pipeline focuses on finding signals that have already merged, it can, in principle, work for signals prior to merger. It does not assume the detector is fixed in time, so it can handle searching over longer stretches of waveform. Specific studies of detection of these sources prior to merger will be needed to more fully understand its capabilities in the pre-merger realm. For pre-merger detection, we expect the full base template search to work. However, its relation to and success of the Heterodyned Likelihood modules is uncertain due to the smaller frequency range over which pre-merger sources will evolve. In principle, this should be okay, but it requires verification. 

When searching for signals after the merger-ringdown, higher-mode waveforms may ensure faster convergence to the true values due to the higher degree of Gaussianity in the higher-mode posteriors. With that said, the peaks in the posterior are thinner, so locating this initial peak may be more difficult than with (2,2)-mode-only waveforms. 

Fast searches are necessary for early detection in low-latency pipelines in order to maximize the potential for simultaneous observation through electromagnetic counterparts. The earlier an MBHB source is detected, the higher the likelihood for coincident detection. The pipeline presented in this work takes $\sim$minutes to locate the maximum Likelihood point associated with the source. This gives an initial point estimate of the sky location. The parameter estimation pipeline can then be run for as long as is necessary to refine the sky-location distribution. As discussed in Section~\ref{sec:segment3}, a roughly accurate map of the extrinsic parameters is attained very quickly during sampling. Therefore, in $\sim10$ minutes, the pipeline could generate posteriors on the sky parameters that are roughly equivalent to the fully converged distributions. A different and optimized method for generating fast sky maps is discussed in \cite{Cornish:2020vtw}.

\section{Conclusion}\label{sec:conclusion}

We presented a solution to the LDC-1A dataset containing a single MBHB signal. The solution involves a fully-automated pipeline that starts with randomized initial points throughout the prior volume. A brute-force Likelihood approach takes advantage of GPU acceleration to make the initial detection of a signal within the data. This portion of the pipeline is flexible and should expand well to signals prior to merger. The brute-force Likelihood can also be directly used for the rest of the pipeline to verify the Heterodyning results or if the Heterodyning technique proves unreliable in any specific situation. The second stage of the pipeline uses a GPU-accelerated batched computation that employs the Heterodyned Likelihood technique. This method is extremely efficient at locating the maximum Likelihood point. Once converged, the final step takes over, running again with the Heterodyning technique to quickly build a final posterior distribution. We provided final posterior distributions for both the noiseless and noise-infused datasets provided by the LDC. Additionally, a posterior was presented for the inclusion of higher-order modes to give the reader a better sense of the true constraints we will be able to place on the signal parameters. For this higher-modes posterior, we injected our own signal with the same parameters as the source from the LDC dataset.

The noisy LDC dataset analyzed in this work was provided with stationary and Gaussian noise properties. The LISA mission will exhibit non-stationary and non-Gaussian noise effects such as from glitches, data gaps, the confusion foreground from Galactic sources, and the generally expected drift of the instrument sensitivity over the observation window. Future datasets produced and provided by the LDC will address these difficult problems. The pipeline designed in this work heavily leverages frequency-domain information. When dealing with the non-stationary information in the future, this pipeline will most likely require a switch to the time-frequency or wavelet domain \cite{Cornish:2020odn}. 

Another realistic complication to be added is the occurrence of multiple simultaneous signals. The LDC has provided a new dataset (LDC-2A) that combines the millions of (mostly unobservable) Galactic binaries with $\sim15$ MBHB sources while maintaining Gaussian and stationary noise assumptions. This dataset is the next logical test for the pipeline designed here. We expect this pipeline to succeed in the search for sources in this new dataset. With the merger-ringdown, where most of the SNR is accumulated, occurring on a timescale of minutes to hours, we expect the $\sim15$ MBHB sources in the data to be sufficiently separated so as to easily distinguish individual sources. The posterior distributions from individually performed parameter estimation runs may be biased in this more-complicated dataset by the presence of other MBHB signals and/or Galactic binary sources. To avoid these biases, a global fit will have to be performed \cite[e.g.][]{Littenberg:2020bxy}. The posterior samples attained from our individual source pipeline may be useful for providing the global fit with direct information on the MBHBs, potentially vastly decreasing the time to convergence of the MBHB portion of the global fit algorithm.

While these more realistic settings require further development of this efficient and automated pipeline, the initial implementation shows promising results indicating we have taken a strong first step towards achieving these long-term LISA analysis objectives.

\acknowledgments

M.L.K. would like to thank Sylvain Marsat, Jonathan Gair, Alvin Chua, Stanislas Babak, Lorenzo Speri, and Ollie Burke for helpful discussions towards the success of this work. This research was supported in part through the computational resources and staff contributions provided for the Quest/Grail high performance computing facility at Northwestern University. This paper also employed use of \texttt{Scipy} \cite{scipy}, \texttt{Numpy} \cite{Numpy}, and \texttt{Matplotlib} \cite{Matplotlib}.


%


\end{document}